\documentclass[pdflatex,sn-mathphys-num]{sn-jnl}

\usepackage{graphicx}%
\usepackage{multirow}%
\usepackage{amsmath,amssymb,amsfonts}%
\usepackage{amsthm}%
\usepackage{mathrsfs}%
\usepackage[title]{appendix}%
\usepackage{xcolor}%
\usepackage{textcomp}%
\usepackage{manyfoot}%
\usepackage{booktabs}%
\usepackage{algorithm}%
\usepackage{algorithmicx}%
\usepackage{algpseudocode}%
\usepackage{listings}%

\theoremstyle{thmstyleone}%
\theoremstyle{thmstyletwo}%

\theoremstyle{thmstylethree}%

\begin{document}

\title[Article Title]{Article Title}


\title{Validating Updated Precession and Nutation Formulations for the Next IERS Conventions}

\author[1]{\fnm{M.} \sur{Karbon}}
\author[2]{\fnm{S.} \sur{B\"ohm}}
\author[3]{\fnm{S.} \sur{Lambert}}
\author[4]{\fnm{T.} \sur{Nilsson}}
\author[1]{\fnm{S.} \sur{Belda}}
\author[5]{\fnm{A.} \sur{Escapa}}
\author[1]{\fnm{J.M.} \sur{Ferr\'andiz}}
\author[7,8]{\fnm{C.L.} \sur{Huang}}
\author[6]{\fnm{J.C.} \sur{Liu}}
\author{\vspace{1mm}\\\text{on behalf of JWG IAG 3.1}}

\affil[1]{\orgdiv{Dep. of Applied Mathematics and Aerospace Engineering}, \orgname{University of Alicante}, \orgaddress{\postcode{03080}, \city{Alicante}, \country{Spain}}}

\affil[2]{\orgdiv{Dep. of Geodesy and Geoinformation}, \orgname{Technische Universität Wien}, \orgaddress{\postcode{1040}, \city{Wien}, \country{Austria}}}

\affil[3]{\orgdiv{SYRTE}, \orgname{Observatoire de Paris}, \orgaddress{\city{Paris}, \country{France}}}

\affil[4]{\orgdiv{Geodetic Infrastructure}, \orgname{Lantmäteriet}, \orgaddress{\city{Gävle}, \country{Sweden}}}

\affil[5]{\orgdiv{Dep. of Aerospace Engineering}, \orgname{University of León}, \orgaddress{\postcode{24071}, \city{León}, \country{Spain}}}

\affil[6]{\orgdiv{School of Astronomy and Space Science}, \orgname{Nanjing University}, \orgaddress{\postcode{210023}, \city{Nanjing}, \country{China}}}

\affil[7]{\orgdiv{Shanghai Astronomical Observatory}, \orgname{Chinese Academy of Sciences}, \orgaddress{\postcode{200030}, \city{Shanghai}, \country{China}}}

\affil[8]{\orgdiv{School of Astronomy and Space Science}, \orgname{University of Chinese Academy of Sciences}, \orgaddress{ \postcode{100049}, \city{Beijing}, \country{China}}}

\abstract{An international effort via the IAU/IAG Joint Working Group on Consistent Improvement of Earth's Rotation Theory (CIERT) aims to update the nutation theory to meet the milliarcsecond accuracy required by GGOS. While a companion article provides the theoretical background to the proposed precession-nutation corrections, this work presents their VLBI-based observational validation. The analysis systematically evaluates the revised models within practical application in VLBI analysis. These results provide critical empirical input for the upcoming integration of consistent, validated Earth rotation models into the next IERS Conventions.}




\maketitle

\section{Introduction}
\label{sec:intro}
Over the last several years, the IAU/IAG Joint Working Group on Consistent Improvement of Earth's Rotation Theory (CIERT) was mandated to enhance Precession and Nutation (PN) theories to meet the milliarcsecond-level accuracy required by contemporary space geodesy and the Global Geodetic Observing System (GGOS). To achieve this, the working group focused on theoretical formulation and observational validation, resolving core limitations in the current standards by implementing a physical model for long-term $J_2$ time variations, introducing non-rigid Earth adjustments to the planetary nutation series, and addressing inconsistencies related to the dynamical ellipticity $H_d$. The companion paper \citep{Ferrandiz2026} comprehensively details these theoretical updates.

\noindent This work focuses directly on the validation and performance evaluation of the corrections derived in \citep{Ferrandiz2026}, by systematically testing these updates within Very Long Baseline Interferometry (VLBI) analysis. This study quantifies their impact on systematic errors and highlights specific periods. These findings form a critical input for the integration of revised models into future reference system standards, which the International Earth Rotation and Reference Systems Service (IERS) is explicitly entrusted to monitor and implement.

\section{Theory and Methodology}
\label{sec:theory_methodology}
The observational validation performed in this study is anchored in the revised physical formulations developed by the CIERT working group and explored in detail by \citep{Ferrandiz2026} and the references therein. Only a basic outline is given here.

\subsection{Theoretical Updates}
\label{subsec:theoretical_updates}
The linear approximation for the $J_2$ rate adopted in the conventional IAU 2006 model introduces a physical inconsistency over multidecadal time spans, causing an unmodeled quadratic acceleration in $dX$ (gray dots in Fig.~\ref{fig:Corr}, right). To mitigate this drift within the initial testing phase, the correction was parameterized strictly as a linear trend (yellow lines in Fig.~\ref{fig:Corr}, left). The advanced quadratic variation formulation designated as $IAU2006\_J2$ \citep{Liu2025}, released in December 2025, is currently undergoing comprehensive analysis (cyan line in Fig.~\ref{fig:Corr}, left); with a preliminary assessment comparing the two approaches presented at the conclusion of this study. To restore physical consistency, a two-fold correction strategy is implemented: introducing semi-analytical non-rigid Oppolzer terms to the planetary nutation components (\citep{Ferrandiz2018}, magenta and black lines in Fig.~\ref{fig:Corr}, left) and applying observation-tuned updates to the dominant lunisolar nutation amplitudes (purple lines in Fig.~\ref{fig:Corr}, left).

\subsection{Free Core Nutation Modeling}
\label{subsec:fcn_modeling}

The Free Core Nutation (FCN) constitutes a primary source of stochastic variance in observed CPO time series, driving resonant responses under tidal forcing that preclude strictly deterministic modeling. Nonetheless, FCN models remain indispensable for CPO prediction and orbit determination. The empirical estimation of the FCN trajectories follows the mathematical framework detailed by \citep{Belda2017}, and three FCN models were determined (depicted in Fig.~\ref{fig:Corr}, center) and made available to the JWG for evaluation: $FCN_{JMF}$ (derived from the ivs24q3 series), and $FCN_{SB1}$ and $FCN_{SB2}$ (derived from IERS C04-20u24, respectively including or omitting offsets). The models consist of daily correction values for dX and dY in $\mu$as covering the timespan 1979.0 - 2026.0.

\subsection{Application of CPO Corrections}
\label{subsec:cpo_corrections}

\begin{figure*}[h]
 \begin{minipage}{1.0\textwidth} 
 \begin{center}
   \begin{minipage}{0.5\textwidth}
        \hspace*{-4ex}{\includegraphics[trim=0cm 0cm 0cm 0cm, clip, width=0.965\textwidth]{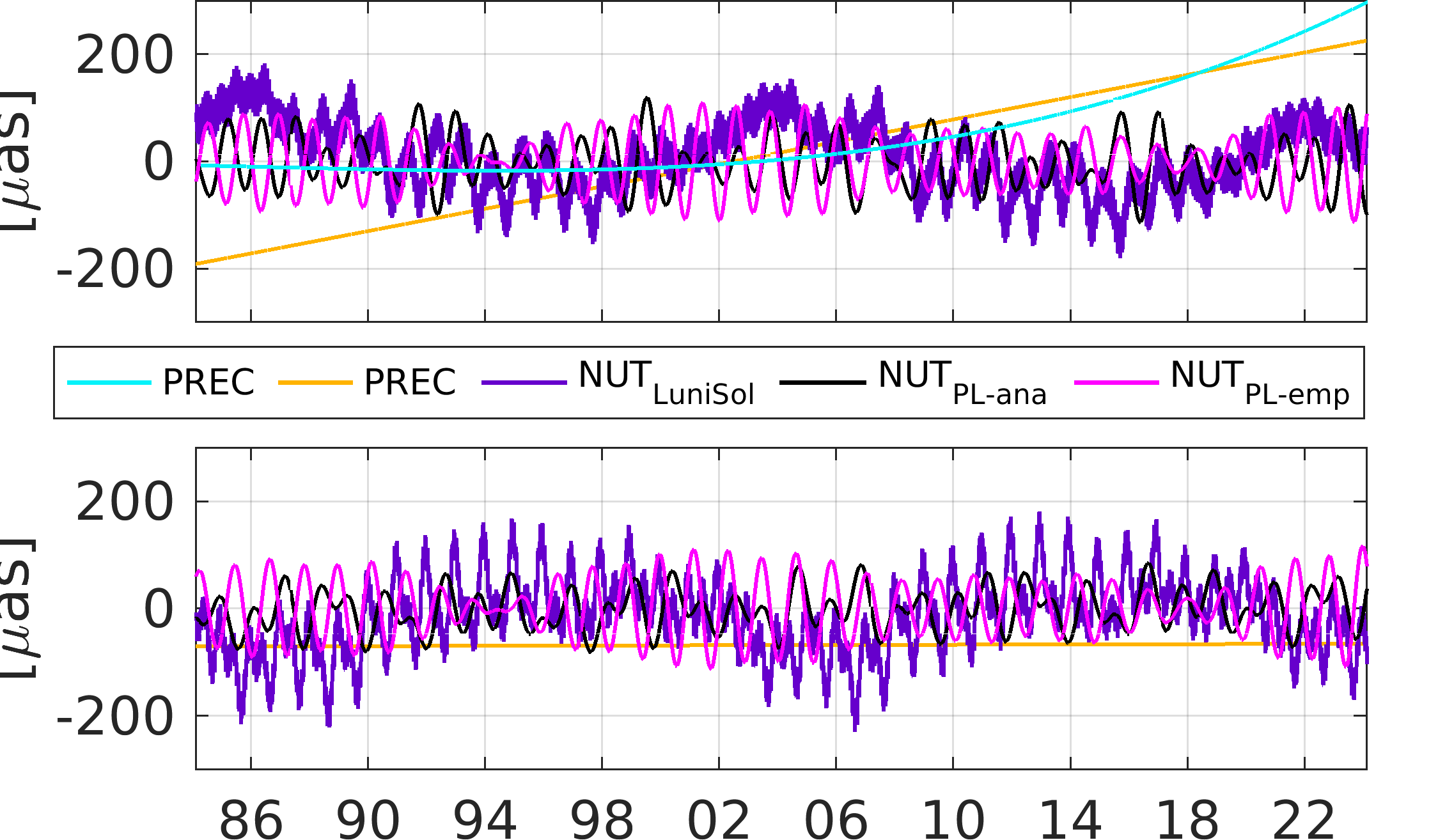}}
   \end{minipage}\hfill
   \begin{minipage}{0.5\textwidth}
        \hspace*{-2ex}{\includegraphics[trim=0cm 0cm 0cm 0cm, clip, width=0.965\textwidth]{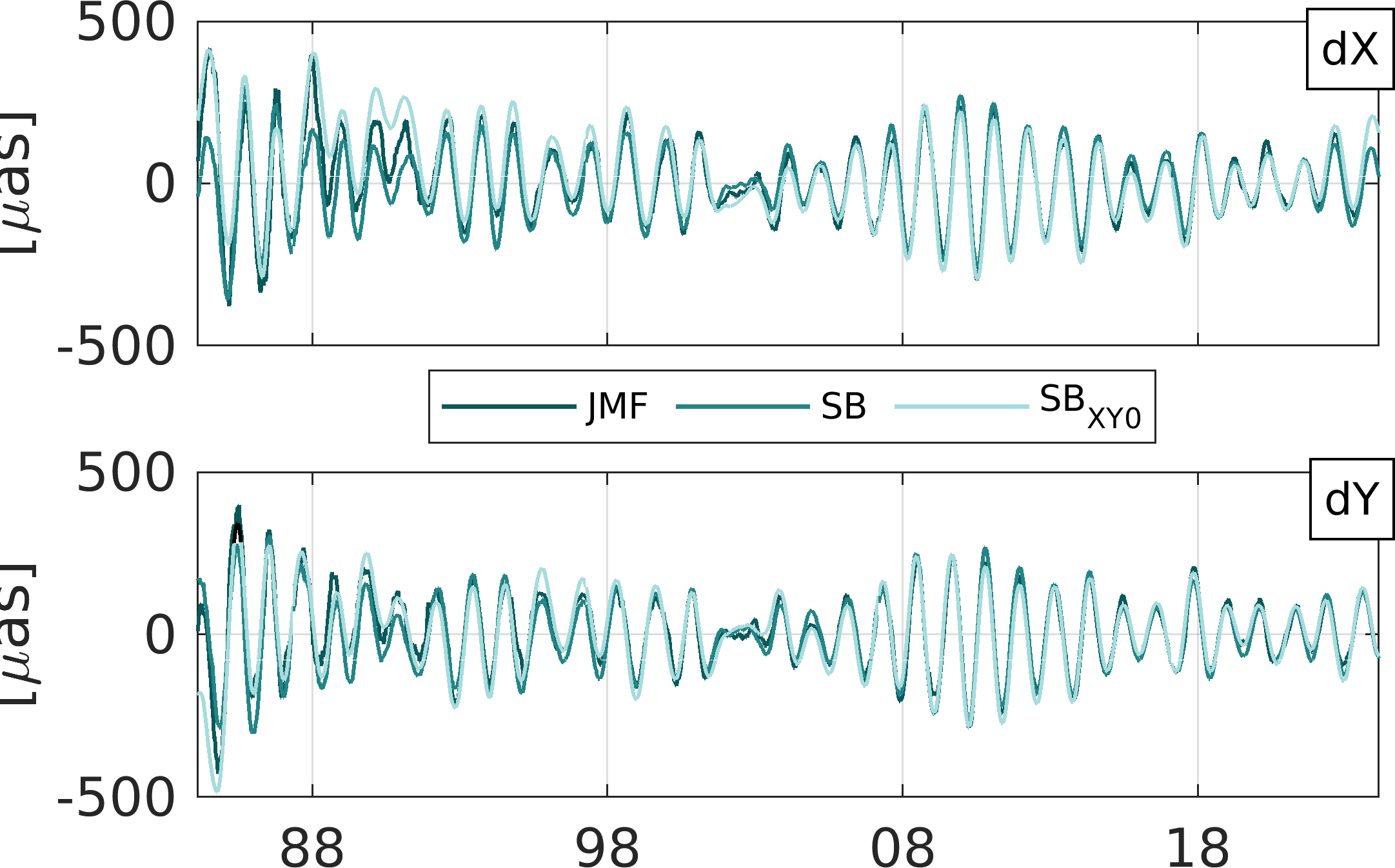}}
   \end{minipage}\hfill
   \begin{minipage}{0.5\textwidth}
        \hspace*{0ex}{\includegraphics[trim=0cm 0cm 0cm 0cm, clip, width=0.965\textwidth]{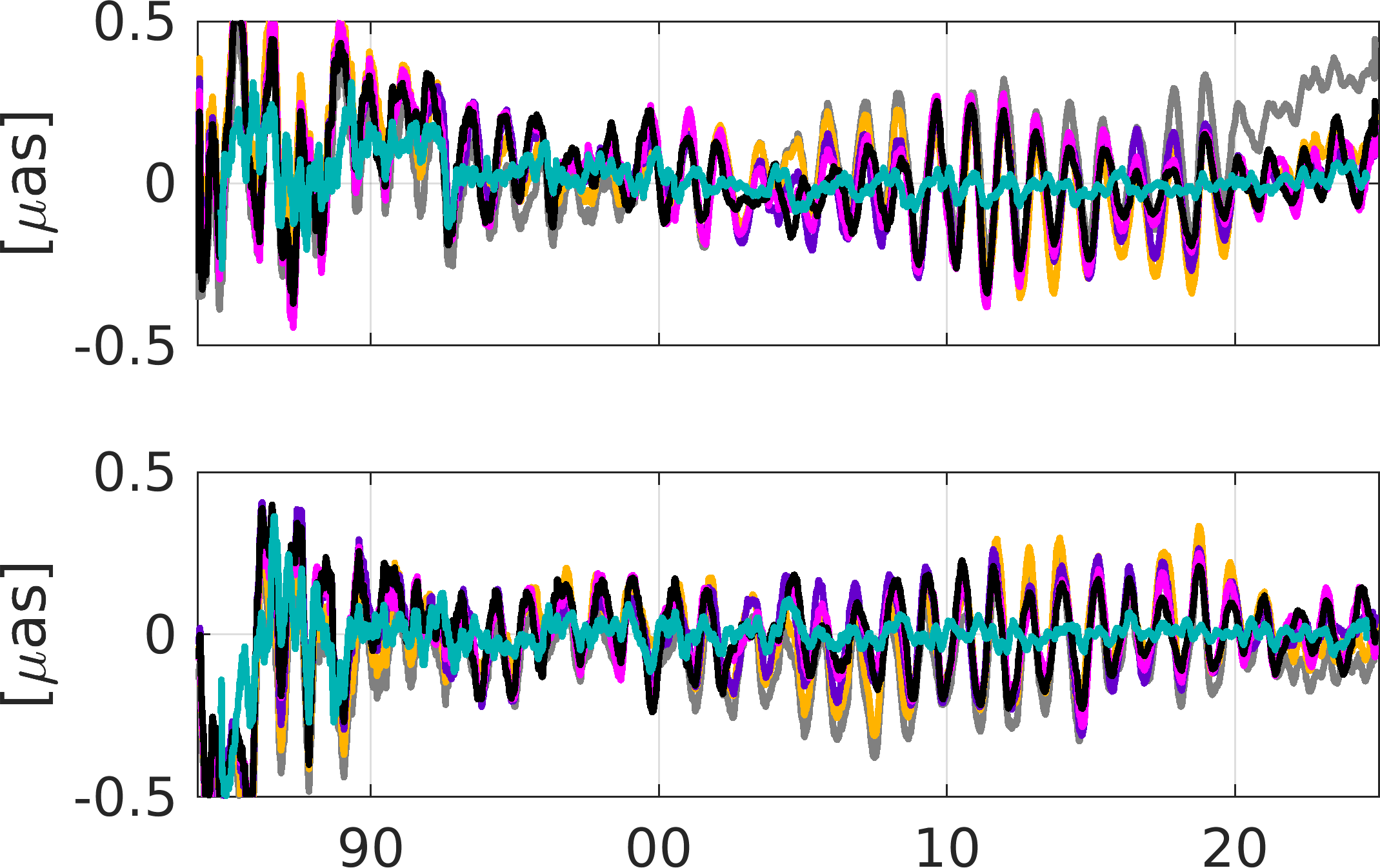}}
   \end{minipage}
 \end{center} 
   \caption{\label{fig:Corr} \textbf{(Upper left)}: Corrections to precession-nutation models split into its various components. Precession: linear trend in yellow, quadratic according to new J2 in green; Nutation: luni-solar cor- rections in purple, analytical planetary in black, and empirical planetary in magenta. \textbf{Upper right}: FCN models. \textbf{Bottom}: CPO after applying each correction to ivs24q3, color-coded accordingly. A sliding window of 90 days was applied to the right panel for visual clarity.} 
 \end{minipage}
\end{figure*}

As detailed in \citep{Ferrandiz2026} the corrections can simply be added to the IAU2000a/2006$_{R06}$ model, or subtracted from CPO time series. The impact of sequentially subtracting the individual precession and nutation corrections from the ivs24q3 reference time series is summarized in Table~\ref{tab:wrms_comparison_final} and visualized in the right panel of Fig.~\ref{fig:Corr}. The plots maintain a consistent color-coding corresponding to each stage: the linear precession trend in yellow, the updated quadratic one in cyan, lunisolar updates in purple, analytical planetary corrections in black, and empirical planetary corrections in magenta.

\noindent When considering the entire data from 1984, the effects of the corrections are severely limited, due to the lower quality of the early VLBI. Focusing on the years 2000 and onward, the improvements are obvious: including the corrections to the IAU2000a/2006$_{R06}$ yields a reduction in $wrms$ of 30$\%$ in $dX$ and of 24$\%$ in $dY$. The inclusion of a FCN model, here exemplarily $FCN_{\text{JMF}}$, yields the most significant reduction in residual scatter, achieving the maximum total cumulative improvements ($\%_{\text{impr}}$) relative to the uncorrected  reference time series ivs24q3. 

\begin{table*}[h]
\centering
\begin{center}
\footnotesize
\setlength{\tabcolsep}{6pt} 
\begin{tabular}{ll ccc cc}
\toprule           
& & \multicolumn{2}{c}{\textbf{Timespan: 1984–2025}} & & \multicolumn{2}{c}{\textbf{Timespan: 2000–2025}} \\
& Correction Stage & $wrms$ [$\mu$as] & $\%_{\text{impr}}$ & & $rms$ [$\mu$as] & $\%_{\text{impr}}$ \\
\midrule    
\multirow{5}*{$dX$} 
& ivs24q3                    & 230.6  & —    & & 198.6 & —       \\
& $-$ PREC                   & 231.8  & -1 & & 166.8 & 16  \\
& $-$ NUT$_{\text{PL-ana}}$  & 221.4  & 4  & & 152.9 & 23 \\
& $-$ NUT$_{\text{PL-emp}}$  & 213.4  & 7  & & 139.1 & 30 \\
& $-$ NUT$_{\text{LuniSol}}$ & 208.0  & 10  & & 139.0 & 30 \\
& $-$ FCN$_{\text{JMF}}$     & 154.8  & 33 & & 87.0  & 56 \\    
\midrule       
\multirow{5}*{$dY$} 
& ivs24q3                    & 236.4 & —    & & 179.1 & —       \\
& $-$ PREC                   & 220.0 & 7  & & 162.8 & 9  \\
& $-$ NUT$_{\text{PL-ana}}$  & 210.4  & 10  & & 148.4 & 17 \\
& $-$ NUT$_{\text{PL-ana}}$  & 208.2 & 12 & & 135.9 & 24 \\
& $-$ NUT$_{\text{LuniSol}}$ & 207.0 & 13 & & 135.5 & 24 \\
& $-$ FCN$_{\text{JMF}}$     & 156.6 & 34 & & 87.0  & 53 \\ 
\bottomrule
\end{tabular} 
\end{center}
\caption{\label{tab:wrms_comparison_final} $rms$ residuals and cumulative improvement in ($\%_{\text{impr}}$) relative to the initial uncorrected ivs24q3 time series, for the long historical horizon (1984–2025) and the modern era (2000–2025).}
\end{table*}

For this study, the individual contributions were aggregated, resulting in a single daily correction term for dX and dY in $\mu$as, respectively, covering the period from 1979.0 to 2026.0:
\begin{equation}
PN_{\text{corr}} = PREC + NUT_{PL-ana} + NUT_{PL-emp} + NUT_{LuniSol}
\label{eq:sum} 
\end{equation}

\subsection{Evaluation Framework}
\label{subsec:evaluation_setup}
In total, one set of corrections to the precession-nutation model was provided ($PN_{corr}$), accumulating the individual components following Eq.~\ref{eq:sum}, alongside three FCN models $(JM F, SB 1, SB 2)$. All corrections are formatted as daily values in $\mu$as for dX and dY, respectively, spanning 1979.0–2026.0.

To validate the operational performance of these corrections and models, an analysis configuration was agreed upon during IAG 2025. The a-priori setup incorporated ICRF3, ITRF2020\textunderscore u2023, and the C04-20u23 EOP series; the latter used exclusively for polar motion and UT1. Standard IVS settings were applied to all remaining parameters. The corresponding session lists (ITRF2020, ITRF2020\textunderscore u2023, and ITRF2020\textunderscore u2024) are accessible via the IVS Analysis Coordinator platform. 

Finally, three analysis centers submitted contributions across three distinct correction setups: uncorrected ($noCorr$), precession-nutation corrections applied (PN), and precession-nutation with an additional FCN model ($FCN_{JMF}, FCN_{SB1}, FCN_{SB2}$):
\begin{itemize}
    \item {Paris Observatory (OPAR):} CALC/SOLVE 7413 sessions, 1984–2025.
    \item {Onsala Space Observatory (ONSA):} ASCOT, 7409 sessions, 1984–2025.
    \item {Vienna University of Technology (VIE):} VieVS 6293 sessions subject to network selection criteria, 1984–2024.
\end{itemize}

\begin{figure*}[htbp]
    \centering
    \begin{minipage}{0.48\textwidth}
        \centering
        \includegraphics[width=\textwidth]{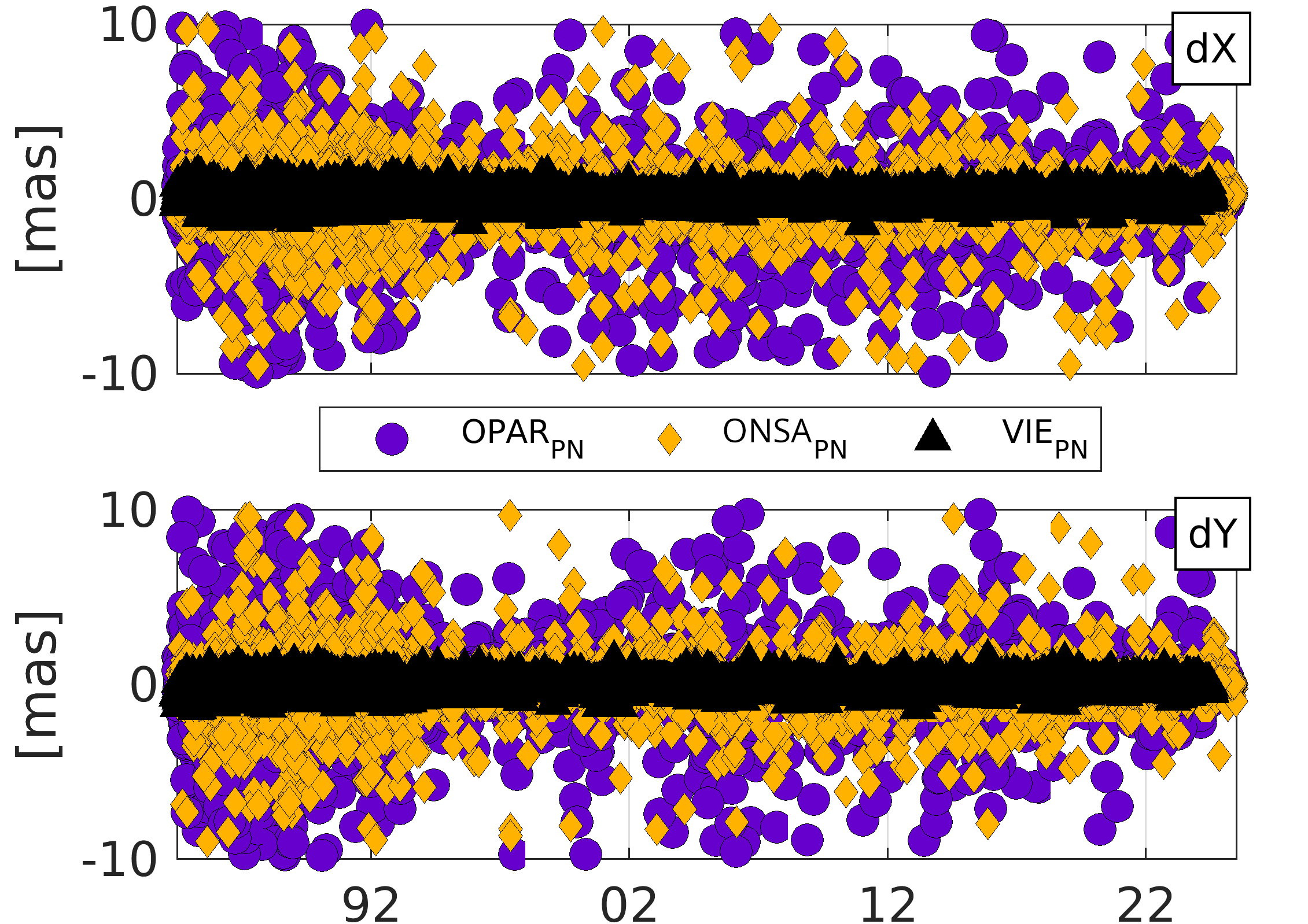}
    \end{minipage}\hfill
    \begin{minipage}{0.48\textwidth}
        \centering
        \includegraphics[width=\textwidth]{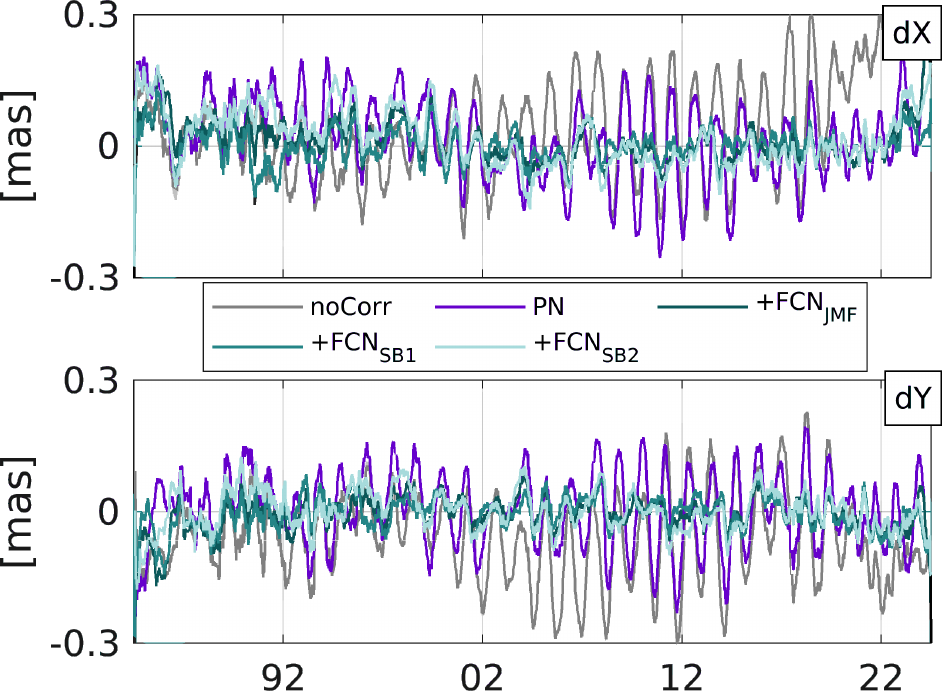}
    \end{minipage}
    \vspace{2mm}
    \caption{\textbf{Left:} CPO residuals derived after correcting precession and nutation (PN) for OPAR (purple), ONSA (yellow), black for VIE. \textbf{Right:} Time series for VIE applying the models cumulatively, in gray: no corrections ($noCorr$), purple: $PNcorr$ applied (PN), 3 shades of teal: FCN model. A sliding window of 90 days was applied to the right panel for visual clarity.}
    \label{fig:combined_cpo_plot}
\end{figure*}


\section{Evaluation Results} 
The results presented hereafter are based exclusively on the shortened data set from the year 2000 onward. While truncating the input data enhances the statistical metrics and optimizes the spectral resolution of shorter periodic components, it has to be kept in mind that it weakens the formal determination of the long-period 9.3-year and 18.6-year nutation terms. 

The left of Fig.~\ref{fig:combined_cpo_plot} shows exemplary the CPO residuals for all ACs for the solution where $PN_{corr}$ is applied, while the right one depicts the progressive reduction of the residuals for VIE, from gray (uncorrected), to purple ($PN_{corr}$ applied) to three shades of teal ($FCN$ applied); for visual clarity a 90 day sliding window was applied.

The solutions show distinctly different noise characteristics: OPAR is the most affected by noise, VIE is remarkably smooth, and ONSA lies in between. This is also reflected throughout the $wrms$ values in Tab.~\ref{tab:wrms-all}, where the first row gives the $wrms$ of the uncorrected time series ($noCorr$), the second bold with $PN_{corr}$ \ref{eq:sum} applied, and then the values for applying additionally the 3 FCN models. The percentages give the cumulative reduction of the $wrms$ in \%. 

As Tab.~\ref{tab:wrms-all} illustrates, applying the corrections to the precession-nutation yields a significant improvement w.r.t. the raw data ($noCorr$), with error reductions ranging from approximately 6\% to 28\%. The most substantial improvements occur when the FCN models are applied on top of the PN corrections. Reductions reaching up to 47.8\% in the VIE solution can be reached, however no FCN model seems to outperform the others. 
In the last column the reference values w.r.t. ivs24q3 from Tab.~\ref{tab:wrms_comparison_final} are given, highlighting the anticipated improvements, and it can be seen, that VIE, which shows exceptional low scatter, almost reaches this threshold.

\begin{table*}[h]
    \centering
    \footnotesize
    \begin{tabular}{lllllllll}
        \toprule      
        & & \multicolumn{2}{c}{\textbf{OPAR}} & \multicolumn{2}{c}{\textbf{ONSA}} & \multicolumn{2}{c}{\textbf{VIE}} & \textbf{ivs24q3$_{corr}$} \\
        &     [$\mu$as] & $wrms$ & \% & $wrms$ & \% &  $wrms$ & \%  \\
        \midrule
        \multirow{5}*{$dY$}&  noCorr & 269.92 & & 226.95 & & 184.96 & & \\
        & \textbf{PN$_{corr}$} & \textbf{234.98} & \textbf{12.9 \%} & \textbf{187.27} & \textbf{17.5 \%} & \textbf{132.34} & \textbf{28.4 \%} &  \textbf{30\%} \\
        & FCN$_{JMF}$ & 222.31 & 17.6 \% & 164.24 & 27.6 \% & 98.86 & 47.6 \% & 56\% \\
        & FCN$_{SB1}$ & 229.10 & 15.1 \% & 164.49 & 27.5 \% & 96.53 & 47.8 \% & \\
        & FCN$_{SB2}$ & 220.77 & 18.2 \% & 167.28 & 26.3 \% & 103.35 & 44.1 \% & \\    
       \midrule        
        \multirow{5}*{$dY$}& noCorr & 261.37 & & 205.11 & & 175.11 & & \\
        & \textbf{PN$_{corr}$} & \textbf{246.20} & \textbf{5.8 \%} & \textbf{180.93} & \textbf{11.8 \%} & \textbf{135.07} & \textbf{22.9 \%} & \textbf{24\%} \\
        & FCN$_{JMF}$ & 228.36 & 12.6 \% & 154.77 & 24.5 \% & 95.65 & 45.4 \% & 51\% \\
        & FCN$_{SB1}$ & 230.3 & 11.9 \% & 155.30 & 24.3 \% & 95.79 & 45.3 \% & \\
        & FCN$_{SB2}$ & 228.40 & 12.6 \% & 156.85 & 23.5 \% & 99.27 & 43.3 \% & \\ 
        \bottomrule
    \end{tabular}
    \caption{$wrms$ and \% of reduction of residuals in dX and dY w.r.t the corrections and models applied. The last column gives the reduction anticipated in Tab.~\ref{tab:wrms_comparison_final}. Based on the time span 2000-2024 and same sub-set of $\sim$3500 sessions.}
    \label{tab:wrms-all}
\end{table*}

\subsection{Spectral analysis} 
\begin{figure*}[htbp]
    \centering
    \begin{minipage}{0.37\textwidth}
        \centering
        \includegraphics[width=\textwidth]{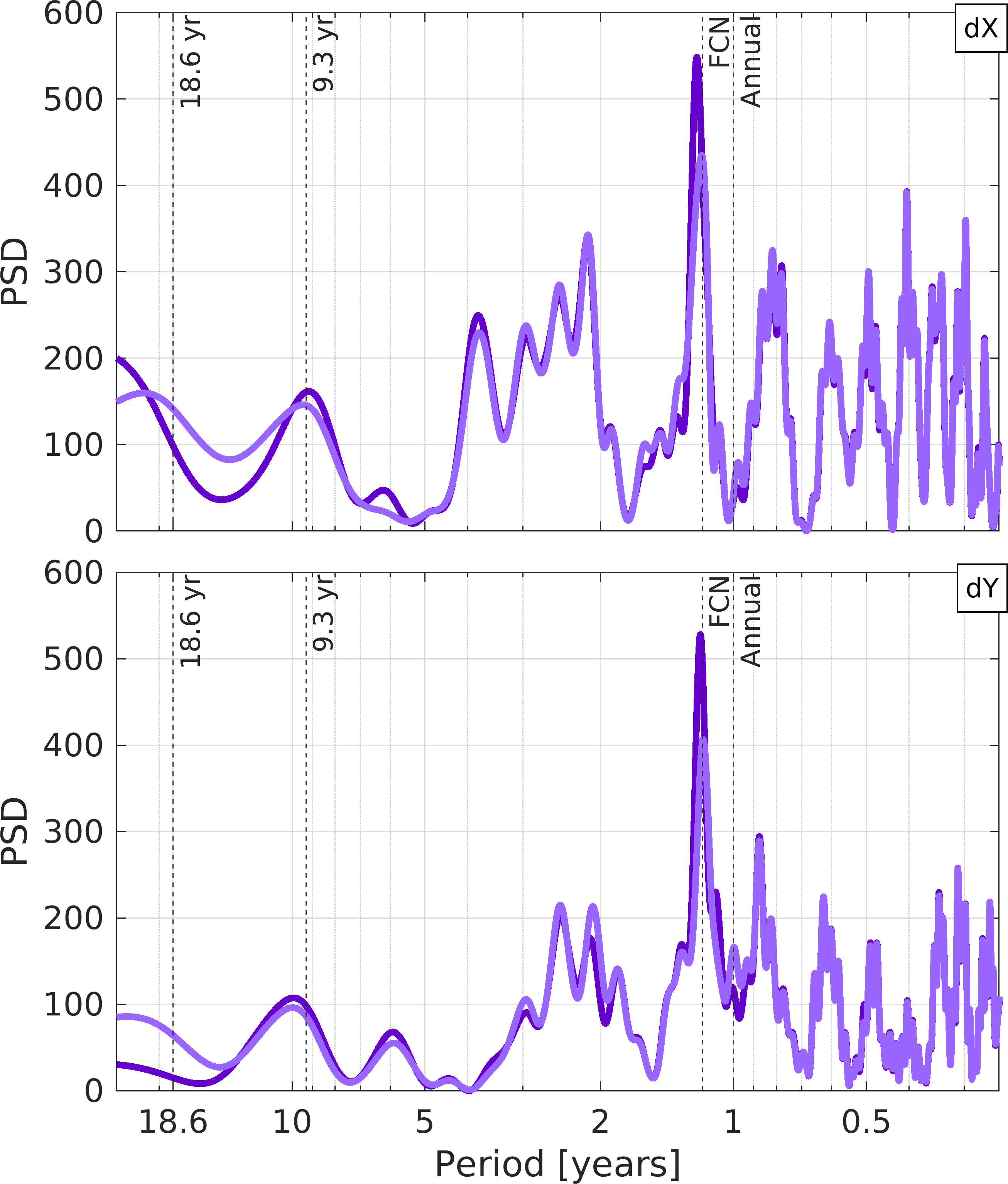}
    \end{minipage}\hfill
    \begin{minipage}{0.63\textwidth}
        \centering
        \includegraphics[width=.95\textwidth]{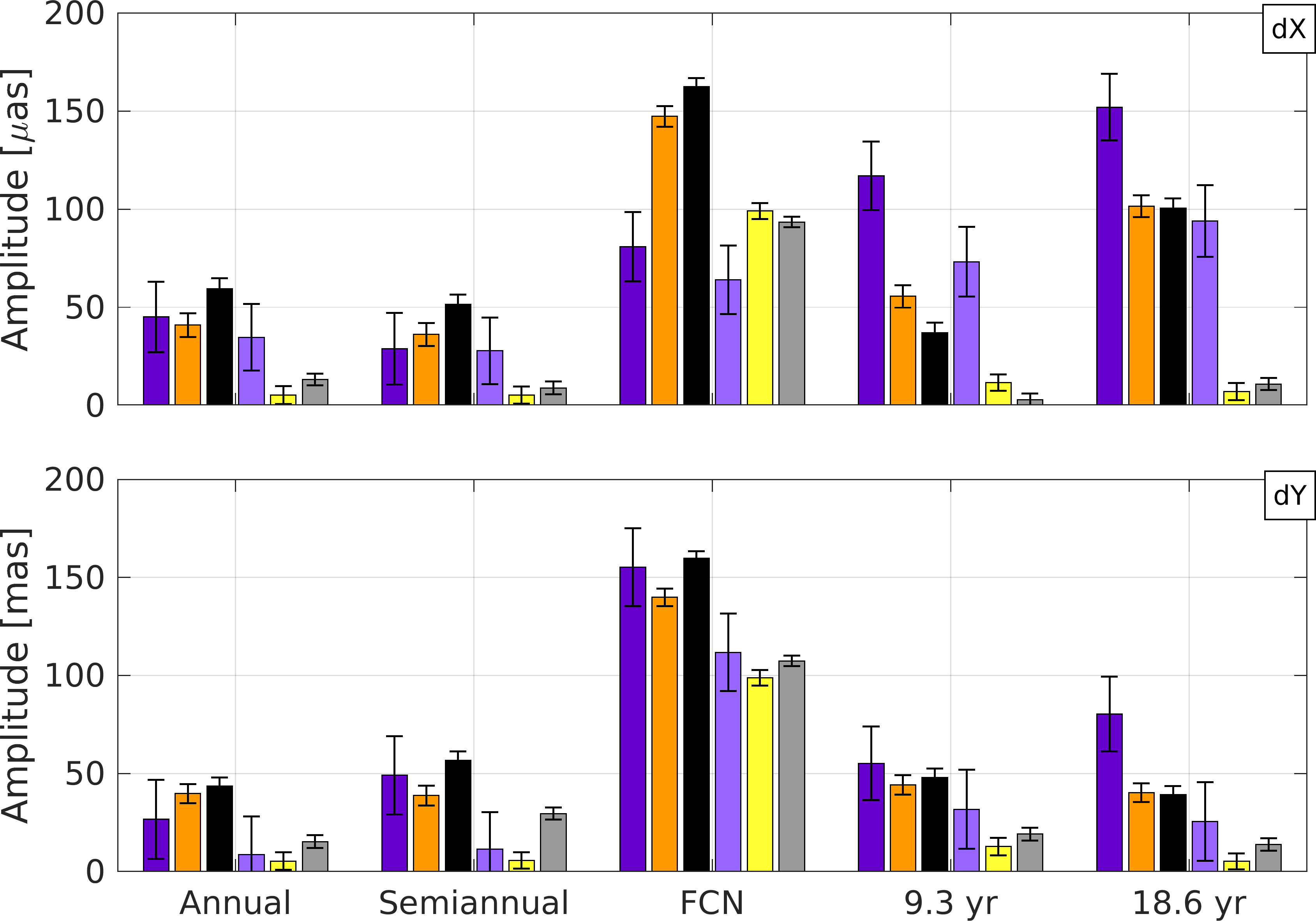}
    \end{minipage}
    \vspace{2mm}
    \caption{\textbf{Left:} PSD of the OPAR CPO, in dark purple the uncorrected signal, in the lighter color, the corrected one. \textbf{Right:} Weighted Harmonic Amplitudes for specific periods, in the full color for the uncorrected time series, in the faded ones for the corrected: OPAR (purple), ONSA (yellow), and VIE (black). }
    \label{fig:PSD-OPAR-all}
\end{figure*}

To further evaluate the characteristics of the precession-nutation corrections, a spectral analysis was performed. The left panel of Fig.~\ref{fig:PSD-OPAR-all} illustrates exemplarily the power spectral density (PSD) of the OPAR CPO time series, with prominent periods marked by vertical lines (FCN @430d). The right panel displays the corresponding amplitudes estimated via a weighted least-squares fit at these periods for all solutions: $noCorr$ are shown in full colors for OPAR, ONSA, and VIE; while their $PN_corr$ counterparts are shown in faded colors. Although all spectral amplitudes decrease after applying the corrections, the magnitude of this reduction reflects the $wrms$ performance in Tab.~\ref{tab:wrms-all}. Specifically, less noisy solutions allow for a clearer separation of frequencies, thereby enabling a more reliable assessment of the model updates' impact. For example, the signature of the FCN is notably weaker in the OPAR solution than in the other two as  it is largely obscured by noise: the Signal-to-Noise Ratios (SNR) in Tab.~\ref{tab:Periods_all} of OPAR for FCN is 6-8 times smaller for $dX$ and 4-6 times smaller for $dY$ compared to ONSA and VIE. Most importantly, while the FCN amplitudes for ONSA and VIE drop by more than 30\% after applying the corrections, their corresponding SNRs do not decrease significantly, hence the FCN signal is preserved.

\begin{table}[htbp]
	\centering	
	\footnotesize
	\setlength{\tabcolsep}{1.6pt} 
	\begin{tabular}{cl rr c rr}
		\toprule
		&& \multicolumn{2}{c}{$dX$ [$\mu$as]} && \multicolumn{2}{c}{$dY$ [$\mu$as]} \\
		\cmidrule{3-4} \cmidrule{6-7}
		  & Period  & \multicolumn{1}{c}{Uncorrected (SNR)} & \multicolumn{1}{c}{Corrected (SNR)} && \multicolumn{1}{c}{Uncorrected (SNR)} & \multicolumn{1}{c}{Corrected (SNR)} \\
		\midrule
		\multirow{5}{*}{\rotatebox{90}{\textbf{OPAR}}} 
		& Annual     &  45.11 \textpm\ 17.97 (2.5) &  34.69 \textpm\ 16.92 (2.1) &&  26.70 \textpm\ 20.15 (1.3) &   8.65 \textpm\ 19.43 (0.5)\\
		& Semia.     &  28.94 \textpm\ 18.24 (1.6) &  27.84 \textpm\ 16.96 (1.6) &&  49.09 \textpm\ 19.96 (2.5) &  11.40 \textpm\ 18.97 (0.6)\\
		& FCN        &  80.86 \textpm\ 17.61 (4.5) &  63.98 \textpm\ 17.37 (3.7) && 155.20 \textpm\ 19.88 (7.8) & 111.79 \textpm\ 19.80 (5.7)\\
		& 9.3 yr     & 116.98 \textpm\ 17.50 (6.7) &  73.20 \textpm\ 17.67 (4.1) &&  55.21 \textpm\ 18.74 (3.0) &  31.76 \textpm\ 20.12 (1.6)\\
		& 18.6 yr    & 151.98 \textpm\ 17.04 (8.9) &  93.94 \textpm\ 18.20 (5.1) &&  80.29 \textpm\ 19.10 (4.2) &  25.48 \textpm\ 20.01 (1.3)\\
		\cmidrule{1-2} \cmidrule{3-4} \cmidrule{6-7}
		\multirow{5}{*}{\rotatebox{90}{\textbf{ONSA}}} 
		& Annual     &  40.96 \textpm\  6.05 (6.8) &   5.33 \textpm\  4.55 (1.2) &&  39.79 \textpm\  4.87 (8.2) &   5.28 \textpm\  4.46 (1.2) \\
		& Semia.     &  36.15 \textpm\  5.83 (6.2) &   5.22 \textpm\  4.37 (1.2) &&  38.78 \textpm\  5.08 (7.6) &   5.64 \textpm\  4.22 (1.3) \\
		& FCN        & 147.23 \textpm\  5.33 (27.6)&  99.02 \textpm\  4.05 (24.5)&& 139.82 \textpm\  4.48 (31.2)&  98.74 \textpm\  3.97 (24.8) \\
		& 9.3 yr     &  55.57 \textpm\  5.72 (9.7) &  11.62 \textpm\  4.24 (2.7) &&  44.24 \textpm\  4.96 (8.9) &  12.75 \textpm\  4.46 (2.9) \\
		& 18.6 yr    & 101.50 \textpm\  5.54 (18.3)&   7.04 \textpm\  4.31 (1.6) &&  40.21 \textpm\  4.76 (8.4) &   5.24 \textpm\  4.09 (1.3) \\
		\cmidrule{1-2} \cmidrule{3-4} \cmidrule{6-7}
		\multirow{5}{*}{\rotatebox{90}{\textbf{VIE}}} 
		& Annual     &  59.47 \textpm\  5.26 (11.3) &  13.24 \textpm\  3.02 (4.4) &&  43.65 \textpm\  4.28 (10.2) &  15.28 \textpm\  3.29 (4.6) \\
		& Semia.     &  51.55 \textpm\  4.99 (10.3) &   8.91 \textpm\  3.25 (2.8) &&  56.77 \textpm\  4.51 (12.6) &  29.55 \textpm\  3.08 (9.6) \\
		& FCN        & 162.38 \textpm\  4.44 (36.5) &  93.40 \textpm\  2.70 (34.5)&& 159.77 \textpm\  3.57 (44.8) & 107.42 \textpm\  2.63 (40.8) \\
		& 9.3 yr     &  37.08 \textpm\  5.04 (7.4)  &   2.88 \textpm\  3.10 (0.9) &&  47.88 \textpm\  4.59 (10.4) &  19.07 \textpm\  3.31 (5.8) \\
		& 18.6 yr    & 100.56 \textpm\  4.87 (20.6) &  10.88 \textpm\  3.11 (3.5) &&  39.25 \textpm\  4.33 (9.1)  &  13.79 \textpm\  3.26 (4.) \\
		\bottomrule
	\end{tabular}
	\caption{Estimated amplitudes for $dX$ and $dY$ for the uncorrected and corrected timeseries for all ACs, in parenthesis the respectiva SNR. All values are in $\mu$as.}
	\label{tab:Periods_all}
\end{table}

\subsection{FCN model evaluation} 

\begin{figure*}[htbp]
    \centering
    \begin{minipage}{0.37\textwidth}
        \centering
        \includegraphics[width=\textwidth]{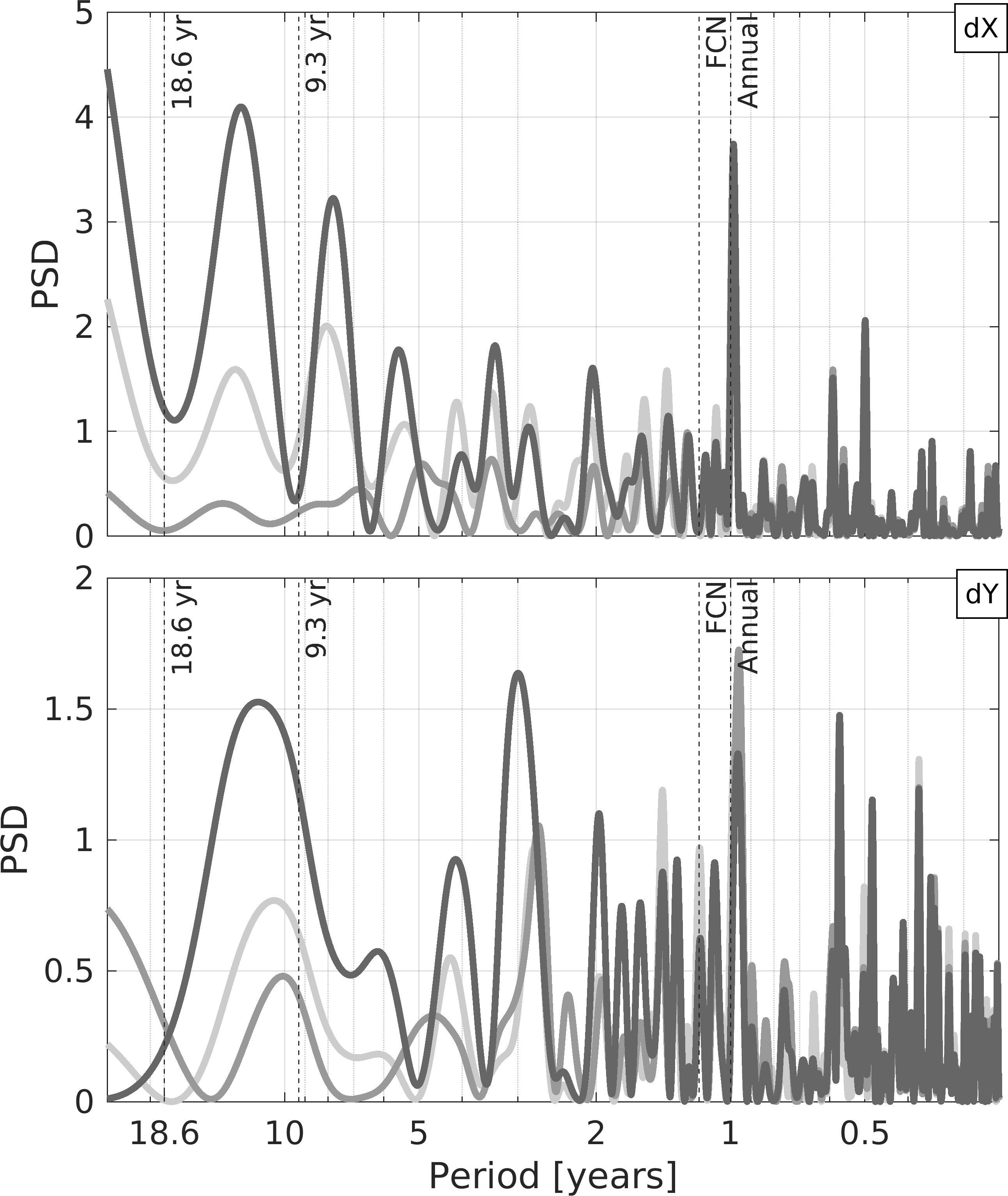}
    \end{minipage}\hfill
    \begin{minipage}{0.63\textwidth}
        \centering
        \includegraphics[width=.95\textwidth]{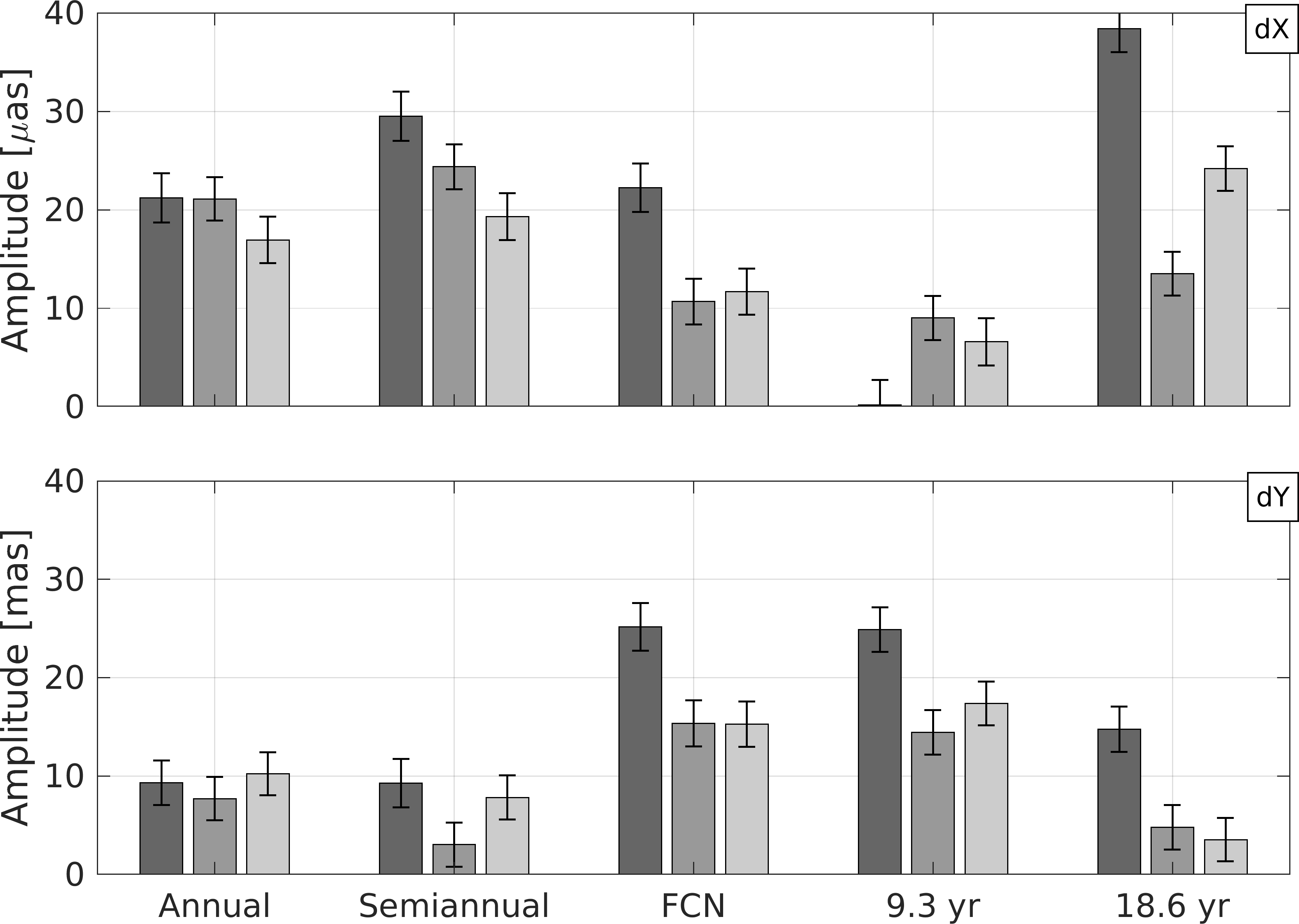}
    \end{minipage}
    \vspace{2mm}
    \caption{\textbf{Left:} PSD of the VIE CPO after correcting for FCN, from the lightest to the darkest: $FCN_{JMF}$, $FCN_{SB1}$, $FCN_{SB2}$. \textbf{Right:} Weighted Harmonic Amplitudes for specific periods.}
    \label{fig:PSD-VIE-FCN}
\end{figure*}

Figure \ref{fig:PSD-VIE-FCN} shows the PSD as well as the amplitudes for the VIE solutions applying additional FCN models (middle panel in Fig.~\ref{fig:Corr}), form the lightest to the darkest gray: $FCN_{JM}$, $FCN_{SB1}$, $FCN_{SB2}$. While in the $wrms$ of time-series no significant differences are visible (Tab.~\ref{tab:wrms-all}), in the PSD as well as in the amplitudes, they become obvious. The $FCN_{SB2}$ seems to be performing worse, as the $SNR$ for the remaining signal in the FCN band is almost double of the other two, further it seems to introduce power to the longest period (18.6 y) in $dX$. This performance gap stems from the omission of the offsets $X_0$ and $Y_0$ in $FCN_{SB2}$. Interestingly, even though the FCN model and its offsets were generated using the C04-20u24 time series instead of ivs24q3, the model including these offsets ($FCN_{SB1}$) still outperforms the alternative. Because all models increase the annual and semiannual amplitudes, the FCN formulation requires further investigation and refinement.

\begin{table}[htbp]
	\centering
	\footnotesize
	\setlength{\tabcolsep}{1.6pt} 
	\begin{tabular}{l ccc c ccc}
		\toprule
		& \multicolumn{3}{c}{$dX$ [$\mu$as]} && \multicolumn{3}{c}{$dY$ [$\mu$as]} \\
		\cmidrule{2-4} \cmidrule{6-8}
		Period & $FCN_{\text{JMF}}$ & $FCN_{\text{SB1}}$ & $FCN_{\text{SB2}}$ && $FCN_{\text{JMF}}$ & $FCN_{\text{SB1}}$ & $FCN_{\text{SB2}}$ \\
		\midrule
		Annual  & $16.9\pm2.4$ (7.2) & $21.1\pm2.2$ (9.5) & $21.2\pm2.5$ (8.5) && $10.2\pm2.2$ (4.7) & $7.7\pm2.2$ (3.5) & $9.3\pm2.3$ (4.1) \\
		Semia.  & $19.3\pm2.4$ (8.1) & $24.4\pm2.3$ (10.7)& $29.5\pm2.5$ (11.8)&& $7.8\pm2.2$ (3.4)  & $3.0\pm2.3$ (1.3) & $9.3\pm2.5$ (3.8) \\
		FCN     & $11.7\pm2.3$ (5.0) & $10.7\pm2.3$ (4.6) & $22.2\pm2.5$ (9.0) && $15.3\pm2.3$ (6.6) & $15.3\pm2.3$ (6.6) & $25.2\pm2.4$ (10.4)\\
		9.3 yr  & $6.6\pm2.4$ (2.7)  & $9.0\pm2.3$ (4.0)  & $0.2\pm2.5$ (0.0)  && $17.4\pm2.2$ (7.9) & $14.4\pm2.3$ (6.4) & $24.9\pm2.3$ (11.0)\\
		18.6 yr & $24.2\pm2.3$ (10.7)& $13.5\pm2.2$ (6.1) & $38.4\pm2.4$ (16.3)&& $3.5\pm2.2$ (1.6)  & $4.8\pm2.3$ (2.1) & $14.8\pm2.3$ (6.4) \\
		\bottomrule
	\end{tabular}
	\caption{Estimated amplitudes for $dX$ and $dY$ components across different FCN empirical model series, their SNR in parenthesis.}
	\label{tab:VIE-FCN}
\end{table}

\subsection{Impact of new $J_2$} 
\begin{figure*}[htbp]
	\centering
	\begin{minipage}{0.37\textwidth}
		\centering
		\includegraphics[width=\textwidth]{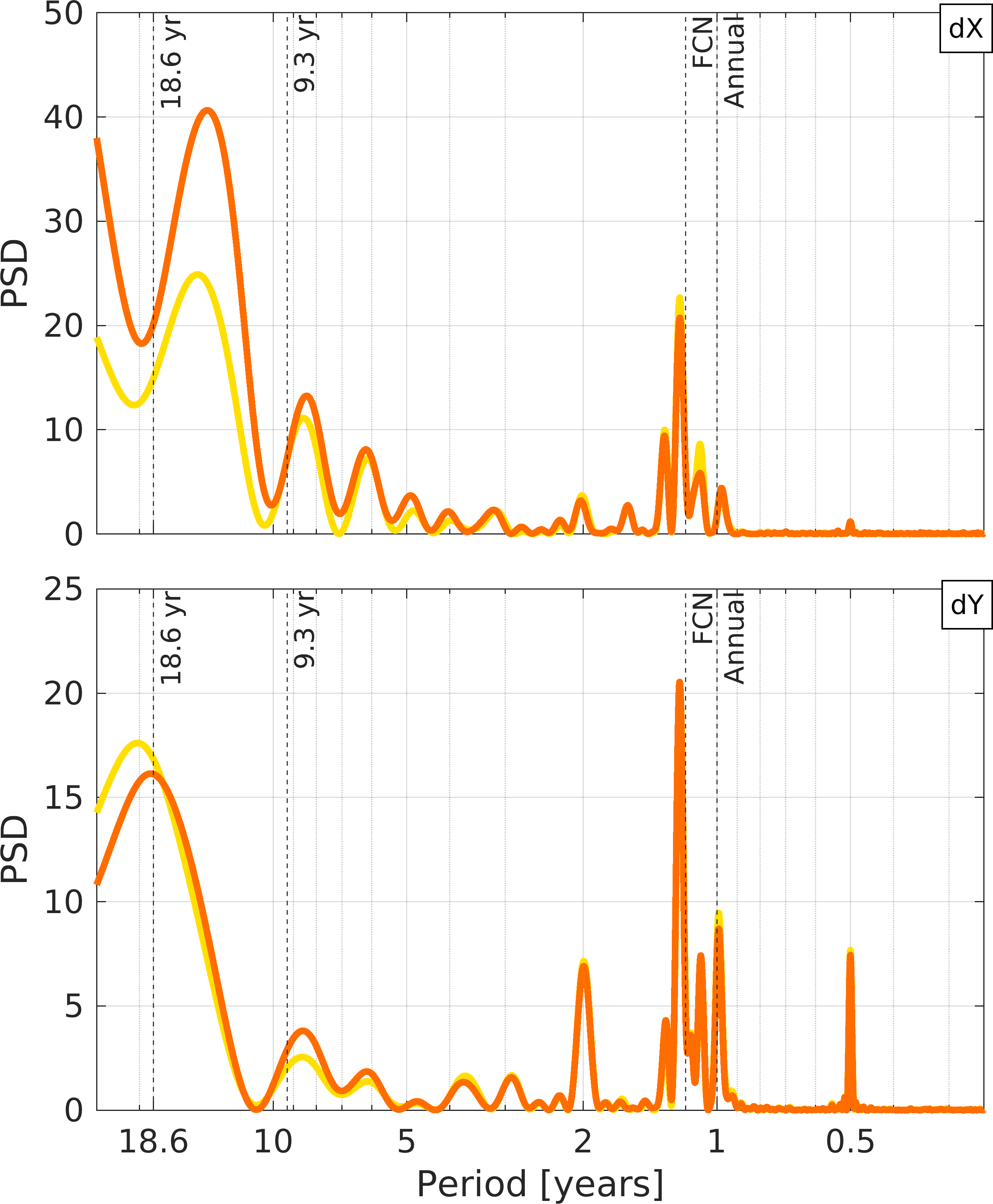}
	\end{minipage}\hfill
	\begin{minipage}{0.63\textwidth}
		\centering
		\includegraphics[width=.95\textwidth]{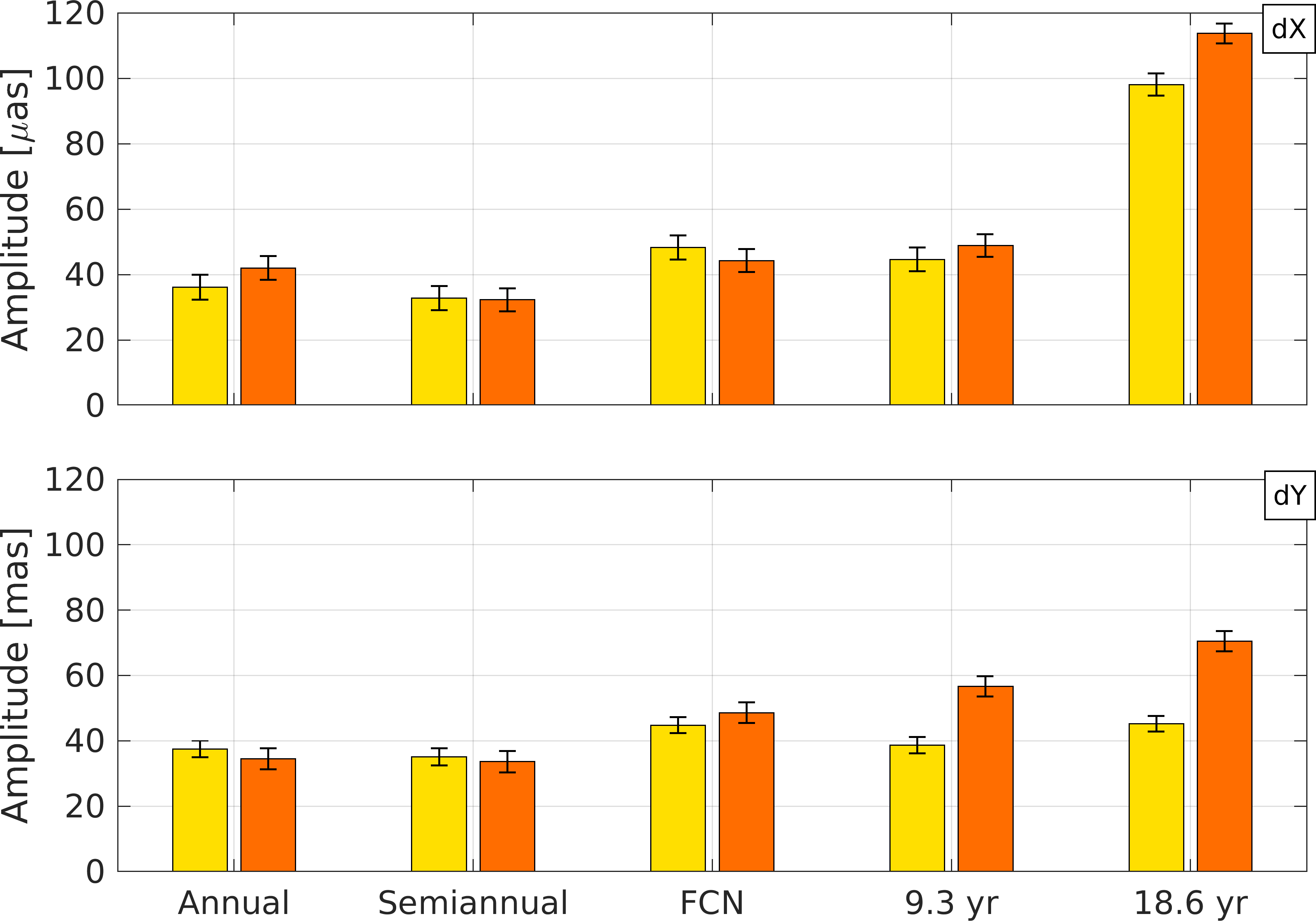}
	\end{minipage}
	\vspace{2mm}
	\caption{\textbf{Left:} PSD of ONSA CPO, in yellow $\Delta linP$ and in orange $\Delta J_2$. \textbf{Right:} Weighted Harmonic Amplitudes for specific periods. }
	\label{fig:J2_PSD_AMP}
\end{figure*}

In Fig.~\ref{fig:J2_PSD_AMP} we show a preliminary result of the updated corrections to the IAU2000a/2006$_{R06}$ model, incorporating the quadratic evolution of $J_2$ (cyan line in left panel in Fig.~\ref{fig:Corr}) based on the ONSA solution. Following the transition from C04-20u23 to finals2000A as the a-priori EOP series, we compare the respective differences, i.e. ($CPO_{noCorr}-CPO_{corr}$), in yellow the solution applying the linear precession correction ($\Delta linP$), in orange the quadratic ($\Delta J_2$). 

As can be seen in the PSD and the estimated amplitudes in Fig.~\ref{fig:J2_PSD_AMP}, the change affects, as expected, mostly the long periods. Interestingly the 18.6yr period in dY shows the largest increase, although the correction is only applied to the dX a-priori. However, the SNR is decreasing significantly for both components, meaning that the statistical significance of the signal drops equally. On the other hand, the SNR for all shorter periods is increased by 30-40$\%$, for the 9.3 yr period by 20$\%$.

\begin{table}[htbp]
	\footnotesize
	\setlength{\tabcolsep}{1.6pt} 
	\begin{tabular}{l rr c rr}
		\toprule
		& \multicolumn{2}{c}{$dX$ [$\mu$as]} && \multicolumn{2}{c}{$dY$ [$\mu$as]} \\
		\cmidrule{2-3} \cmidrule{5-6}
		Period / Term & \multicolumn{1}{c}{$\Delta linP$ (SNR)} & \multicolumn{1}{c}{$\Delta J_2$ (SNR)} && \multicolumn{1}{c}{$\Delta linP$ (SNR)} & \multicolumn{1}{c}{$\Delta J_2$ (SNR)} \\
		\midrule
		Annual     & 36.12 \textpm\ 3.85~~(9.4)  &  41.99 \textpm\ 3.61 (11.6) && 37.46 \textpm\ 2.46 (15.2) & 34.54 \textpm\ 3.22 (17.8)\\
		Semiannual & 32.82 \textpm\ 3.72~~(8.8)  &  32.26 \textpm\ 3.52~~(9.7) && 35.09 \textpm\ 2.60 (13.5) & 33.60 \textpm\ 3.31 (15.7)\\
		FCN        & 48.25 \textpm\ 3.67 (13.1) &  44.26 \textpm\ 3.49 (12.7) && 44.80 \textpm\ 2.48 (18.1) & 48.61 \textpm\ 3.18 (18.8)\\
		9.3 yr     & 44.64 \textpm\ 3.66 (12.2) &  48.87 \textpm\ 3.45 (14.2) && 38.65 \textpm\ 2.47 (15.7) & 56.65 \textpm\ 3.09 (17.6)\\
		18.6 yr    & 98.09 \textpm\ 3.35 (29.3) & 113.73 \textpm\ 3.04 (37.4) && 45.21 \textpm\ 2.39 (19.0) & 70.50 \textpm\ 3.07 (18.1)\\
		\bottomrule
	\end{tabular}
	\centering
	\caption{Amplitudes for $dX$ and $dY$ components between $\Delta linP$ and $\Delta J_2$ together with the SNR in parenthesis for ONSA time series. All values are in $\mu$as.}
	\label{tab:onsa_c04_comparison}
\end{table}

\section{Conclusion}
\label{sec:conclusion}
This report presents a preliminary assessment of the newly updated precession and nutation formulations, with the analysis truncated in scope and detail to satisfy strict manuscript length constraints. The empirical validation confirms that an application of the proposed corrections systematically mitigates systematic errors and residual scatter within modern VLBI time series. A significantly more comprehensive, in-depth investigation including the ongoing refinements to the quadratic $J_2$ formulations is currently in preparation.

\bmhead{Acknowledgements}
This work was supported by the project PID2025-171760NB- I00, funded by MICIU/AEI/10.13039/501100011033 and FEDER, EU; as well as CIEMGE/2026/15. C.L.H. is supported by the National Natural Science Foundation of China (Nos. 12233010), JCL is funded by the National Natural Science Foundation of China (NSFC) under grant No. 12373074.


\bibliography{mybib} 

\end{document}